# DISH-TREND: INTERVENTION MODELING SIMULATOR THAT ACCOUNTS FOR TREND INFLUENCES

Stefan Andjelkovic

Computational and Systems Biology
Joint Carnegie Mellon University and University of Pittsburgh Computational Biology PhD Program
University of Pittsburgh
4420 Bayard St
Pittsburgh, PA 15213, USA

Natasa Miskov-Zivanov

Electrical and Computer Engineering,
Bioengineering, Computational and Systems Biology
University of Pittsburgh
4420 Bayard St
Pittsburgh, PA 15213, USA

## ABSTRACT

Simulation on directed graphs is an important method for understanding the dynamics in the systems where connectivity graphs contain cycles. Discrete Stochastic Heterogeneous Simulator (DiSH) is one of the simulation tools with wide application, which uses regulator values to calculate state updates of regulated elements. Here we present a new simulation approach DiSH-trend which also takes into account the trends in regulating elements. We demonstrate the features of trend-based regulation, as well as hybrid regulation, which is a combination of the trend- and level-based approaches. The modeling capabilities are demonstrated on a small toy model, showcasing different functionalities. Real-world capabilities are demonstrated on a larger network model of food insecurity in the Ethiopian region Oromia. Adding trend-based regulation to models results in increased modeling flexibility, and hybrid regulation improves qualitative dynamic behavior prediction. With appropriate data, DiSH-trend becomes a powerful tool for exploring intervention strategies.

## 1 INTRODUCTION

Complex systems have been studied with network models for the last 50 years. These models encode the system connectivity as the network topology, where system modules are represented as nodes and their interactions as edges. The differences between the models arise from diverse approaches in modeling dynamics. Signal propagation was studied with network models in various fields: biology, psychology, sociology, economics, politics, and engineering (Barrat et al 2004). Modeling the dynamics of the signal propagation helps in understanding how different collective decisions are made, elucidates emergent behaviors and patterns, and offers insight into the effectiveness of potential interventions in solving complex system challenges.

Some of the good examples with the successful application of dynamic networks models are complex biological systems. The networks are composed of interactions between entities such as molecules in cell signaling pathways, or species in an ecosystem. The success of biological network models can be credited with the availability of structured data for parameter tuning. There are several common dynamics modeling approaches: Boolean networks (Albert et al 2008), Discrete networks (Sayed et al 2017), Bayesian networks (Needham et al 2007, Wilkinson 2007), ODE-based models (Materi and Wishart 2007), rule-based models (Faeder et al 2005), and Petri nets (Chaouiya 2007). These approaches vary in flexibility, reliance on data availability, computational expenses, and precision, from the most robust (Boolean networks) to the most precise (ODE-based models).

Here, we are focusing on discrete networks with more than two discrete levels, as a balanced modeling approach that can operate in data-scarce scenarios, and has a tunable resolution to meet modelers' goals.

As an example, we are using the discrete stochastic heterogeneous simulator (DiSH, Sayed et al 2017). This tool simulates signal propagation through a directed graph, by calculating individual element updates according to the states of their regulators (elements with directed edges pointing to the observed element).

DiSH calculates element updates exclusively based on their regulator values. This regulation regime was developed to represent the dynamics of chemical reactions, where a molecular species concentration only increases at the expense of its reagents. While it is a faithful representation of many processes in molecular biology, it lacks the necessary flexibility for models with heterogeneous granularity levels, where some interactions are so "tight" that the regulated element follows the trend of its regulator. This need arises when an interaction is hierarchical in nature so that an increase in the regulator is followed by an increase in the regulated element, and a decrease in the regulator followed by a decrease in the regulated element. As an example, if a population of bees is the only regulator of a population of insects in the model, then population trends in bees should be immediately observed in the population of insects because bees are taxonomically under insects.

In 1957, Holt extended the time series forecasting model in finance where the future levels are estimated as the average of the previous levels, by adding a trend-based term, where the trend is averaged displacement of the time series level over time. (Chatfield 1978, Holt 2004) Holt-Winters model (Winters 1960) expanded on it by introducing seasonality term to account for different seasons, and how certain trends can be damped in certain seasons (e.g., lower productivity increase in winter than summer).

Inspired by the works of Holt and Winters (Holt 2004, Winters 1960) and their persevering efficiency, we present here DiSH-trend, a novel modeling and simulation tool, which accounts for regulator trends alongside their levels and can integrate seasonality through an additional "season" node. The goal is to allow modelers enough flexibility to define the nodes according to both the relevant causal relations and the available data. As an example, assume only bee population data are available, and not insect population in general. If insect population is needed for interactions with other parts of the network, and its data are missing, then bees and insects can be two separate nodes in the network, instead of a single aggregated one called "insect population". In that case, bees, as part of the insect population, will have their population growth trends directly reflected in the total insect population growth.

The synergy of causality (network topology), dynamics modeling (update functions and simulation schemes), and data representation (discretized element values and value toggling) make DiSH-trend a powerful simulation tool for exploring various interventions strategies in complex systems. In this paper we provide a brief overview of discrete network modeling in Section 2, introduce DiSH-trend methods in Section 3. In Section 4, we describe the experiments for testing DiSH-trend capabilities, while in Section 5, we demonstrate qualitative advantages of allowing trend-based regulation in DiSH-trend. Section 6 concludes this paper, briefly summarizing its contributions.

## 2   BACKGROUND

### 2.1   Discrete element-based models

An underlying structure of models that can be simulated by DiSH-trend is a directed graph $G(V,E)$ where each node $v \in V$ represents a single model element, and each directed edge $e \in E$ represents a directed interaction between elements. Elements are storing discrete level variables assigned to the graph nodes, update functions that determine how the element values change over time, as well as additional properties that the modeler finds relevant for human inspection and better model interpretability (e.g., location, category, or the origin of the element). The variable levels are uniformly distributed over [0,1] interval and each element $X_i$ can have its own number of discrete levels $L_i$. For an element with $L$ levels, $k$-th level (where $0 \le k < L$) will have a value $\frac{k}{L-1}$. Element subsets that affect the updates of an element $X_i$ are called influence sets, and the elements that belong to the influence set of the element $X_i$ are called regulators of $X_i$, while $X_i$ is called a regulated element. To simulate dynamics, regulated element values are updated by calculating their update functions. The element update functions are functions of the previous states of their influence sets, and they define the model dynamics.

## 2.2 Simulation schemes

To support different rates and order of changes in system components, DiSH-trend has several different simulation schemes, similar to the schemes implemented in DiSH [Sayed et al 2017]. These simulation schemes are categorized into (i) simultaneous, where all elements are updated simultaneously, (ii) sequential, where elements are updated one-by-one, either following a predefined ordering, or by randomly selecting an element for update in each time step, and (iii) group-update schemes, where each element is assigned to a group, and group elements are updated together. Selection of an arbitrary element/group for updates emulates the stochastic nature of the modeled complex system and allows for sampling over evolution trajectories, by running the simulation multiple times. Assuming no simulated trajectory is a priori more likely than any other, the trajectory is averaged across the runs, as described previously in [Sayed et al 2017]: for each element and each time step, an average value is calculated across all simulation runs, and the sequence of averaged element values is the expected trajectory.

## 2.3 Value toggling

The *input signals* for the discrete network models are time series of element values on the *input nodes* (i.e. nodes with no regulators). They can be encoded through value toggling functionality of the simulator. Similar to DiSH, the new DiSH-trend simulator supports overwriting element values at predefined time steps, by designating value toggle sequence which consists of pairs $(x, t)$, where $x$ is the overwriting value, and $t$ is the time step in which the element value is overwritten. This feature can be used not only for defining input signal, but also for representing shocks to the system, and interventions, by instantaneously changing an element's value, and observing the downstream effects. Leveraging this functionality, a modeler can define relevant scenarios to explore, and compare the outcomes of different intervention strategies. In addition to input nodes, we are denoting nodes with the relevant simulation outcomes as *output nodes*.

## 3 METHODOLOGY

As mentioned in Section 1, the DiSH simulator supported only models where update functions include level-based regulation, that is, element updates are computed solely based on the last observed values of its regulators. Here, we first provide the details of the level-based approach and then introduce the new trend-based modeling approach, the DiSH-trend simulator, which takes into account the trend of changes in regulator values. We will refer to these changes in element values as *trends*, to emphasize that the regulated element only sees the net change in the regulator value between its two consecutive updates. Besides the trend-based regulation, DiSH-trend also supports the hybrid regulation type, a combination of level-based and trend-based regulation, which is detailed at the end of this section.

## 3.1 Level-based regulation

In the level-based regulation approach, each element $X$ is updated according to the function of only the values of its regulator. If $X$ has $n$ regulators, and their values are $R_1, R_2, \ldots, R_n$, then the new value of $X$, $X_{new}$, is computed using the old value of $X$, $X_{old}$, and a function of regulator values as:

$$X_{new} = X_{old} + F(R_1, R_2, \ldots, R_n) \tag{1}$$

Let a regulated element $X$ be regulated by $k$ positive regulator sets and $l$ negative regulator sets, each containing one or more regulators. A regulator set with the regulated element forms a regulation hyperedge (a generalized edge that can have multiple heads and/or tails), with the regulators from a regulator set as the hyperedge tails, and the regulated element as the hyperedge head. We will denote each positive regulator set $\mathbb{I}_{p,i}$, and each negative regulator set $\mathbb{I}_{n,j}$. Contributions from regulators belonging to a common regulator set are multiplied (representing a synergistic effect of these interactions), and then these products are summed (as regulator sets independently regulate the regulated element). If we denote positive regulator terms as $A_i$, negative regulator terms as $I_j$, the corresponding interaction weights as $w_i^{(p)}$, and $w_i^{(p)}$, respectively, the update function $F$ for element $X$ with $L$ discrete levels will have value:

$$F(A_1, A_2, \ldots, A_k; I_1, I_2, \ldots, I_l) = \frac{[L \cdot |B(A,I)|]}{L} \cdot \text{sign}(B(A,I)) \tag{2}$$

where $B(A, I)$ is the balancing function determined by:

$$B(A, I) = \sum_{i=1}^{k} \left( \prod_{q \in \mathbb{I}_{p,i}} w_{i,q}^{(p)} A_{i,q} \right) - \sum_{j=1}^{l} \left( \prod_{r \in \mathbb{I}_{n,j}} w_{j,r}^{(n)} I_{j,r} \right) \tag{3}$$

The products are calculated over the sets of regulators involved in a single regulation hyperedge.

In addition to rounding up to the nearest higher discrete level, the simulator constrains updates to have $0 \leq X_{new} \leq 1$. Any overflow is discarded, as DiSH-trend supports a finite number of states. While this limitation challenges the reliability of the modeled dynamics, note that DiSH-trend supports an arbitrary number of discrete levels for each element, and saturation at maximum level can be interpreted as a suggestion to re-run the simulation with more discrete levels for the saturated element above the reached maximum value. As an example, if $X$ has 6 discrete levels (0, 0.2, 0.4, 0.6, 0.8, and 1), but during the simulation saturates at level 1, modeler can re-run the simulation with 11 discrete levels (0, 0.1, 0.2, …, 0.9, 1). If in the first simulation, the maximum value corresponding to level 1 was $M$, in the following simulation, level 1 can represent value $2M$. This way, simulation span can be iteratively doubled arbitrarily many times, and because the simulation is run for a finite number of timesteps, encapsulating all predicted values of element $X$ during the course of the experiment is warranted. Note that, halving the interval between discrete levels would require halving the weights for consistency, which in turn ensures that the element does not reach the maximum state as fast as in the previous simulation. The same applies for saturation at the minimum level, whereas the extension doubles the value span downwards.

### 3.2 Trend-based regulation

The trend-based regulation extends the update function to include the changes in the regulators, i.e., their trends, as arguments. If only trend-based regulation is present, balancing function for element $X$ is:

$$B_{trend,X}(A, I) = \sum_{i=1}^{k} \left( \prod_{q \in \mathbb{I}_{p,i}} w_{i,q}^{(p)} \Delta A_{i,q} \right) - \sum_{j=1}^{l} \left( \prod_{r \in \mathbb{I}_{n,j}} w_{j,r}^{(n)} \Delta I_{j,r} \right) \tag{4}$$

The balancing function for trend-based regulation is used in the random step-based sequential scheme (RSB-Sequential) [Sayed et al 2017], where the timeline is non-linear due to the stochastic selection of the updated element. In this simulation scheme, which was described in detail in [Sayed et al 2017], a pool of updateable elements (those with regulators) is formed, and in each time step, a single element is randomly picked from this pool to be updated. With this in mind, there are multiple ways one can define the trends. To enforce consistency, and prevent a single regulator update from being observed multiple times in the regulated element, we defined the regulator trends (e.g., for a regulator $R$ of an element $X$) as the difference between the current regulator value (i.e., $R(t)$) and its value at the time step of the previous update of this regulated element $X$ (i.e., $R(t_{last\_update\_of\_X})$):

$$\Delta R_X(t, t_{last\_update\_of\_X}) = R(t) - R(t_{last\_update\_of\_X}) \tag{5}$$

We also updated equation (3) from the previous subsection to account for this change in $R$:

$$X_{new} = X_{old} + F(\mathbf{R}(t), \Delta \mathbf{R}_X(t, t_{last\_update\_of\_X})) \tag{6}$$

This implementation means that each element (except for the input elements, that have no regulators), stores the last two values of all of its regulators. We initialize these doublets of previous values by storing the initial value of the regulator in both slots, that is, its ultimate and the penultimate observed values. Hence, all trends are initialized at 0. While the current version of DiSH-trend supports any initial value of an element (within the set of its allowed values), it does not support nonzero initial trend values. As our next step, we will add a capability to DiSH-trend to use historical data from time before the simulation start date, and thus, to initialize non-zero trends where applicable. During the simulation runs, whenever the regulated elements are updated, their regulators' previous two observed values get updated accordingly. Fig. 1 shows an example of how trend-based calculations are done on a small 3-element model where all interactions are trend-based.

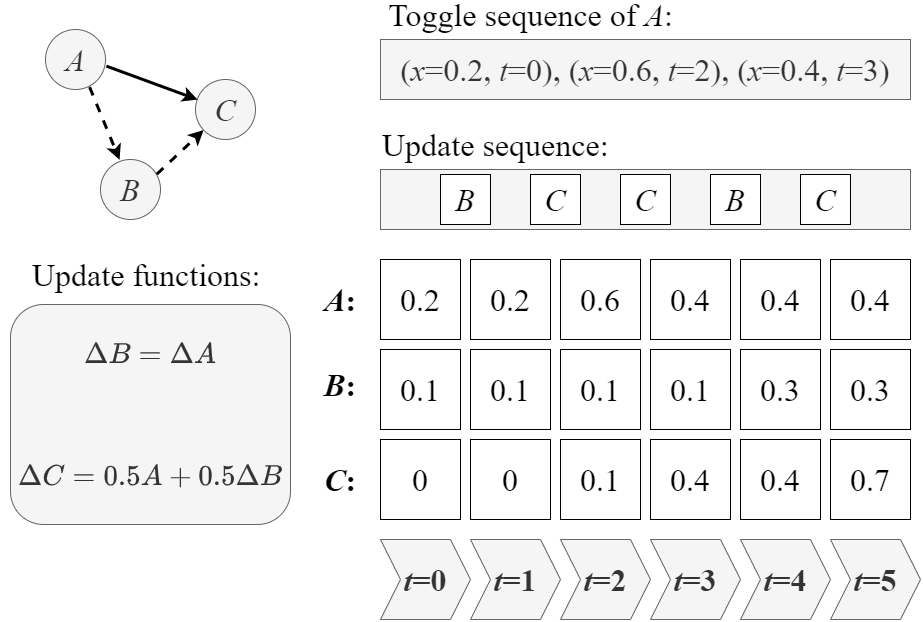

Figure 1: Trend-based calculation demonstrated on 3-element toy model: underlying graph with the full line representing level-based regulation, and dashed line representing trend-based regulation (top left), update functions (bottom left), an input signal on node $A$ (top right), arbitrary update order sequence (right), and the values of elements in each time step (below update sequence). In the first time step, $B$ is selected to be updated, but the trend of $A$ is equal to 0, so $B$ remains unchanged. Then, $C$ is selected and it is increased due to nonzero value of $A$. In time step $t = 3$, $C$ is updated again, proportional to the value of $A$. Element $B$ is updated in time step $t = 4$ and it finally observes the trend of $A$ by comparing its value at time step $t = 3$ and the last time step when $B$ was selected to be updated (time step $t = 0$).

## 4  EXPERIMENTS

We tested the modeling capabilities of our new DiSH-trend simulator on two graphs of different scale. First series of experiments is run on small 4-element synthetic graphs (Fig. 2) to ensure validity of individual features, as well as to compare with the previous simulator version. Second series of experiments was run on larger graph representing a real-world model of food insecurity in Ethiopia. In the larger model we are showing how trend-based regulation can be used on real data and how it gives modelers more flexibility.

The simulations are run on MacBook Pro with 3.3 GHz Dual-Core Intel Core i5, 16 GB 2133 MHz LPDDR3 memory. The simulator is implemented in Python 3.8.5 and is available at https://bitbucket.org/andjelkovicstefanpitt/dish-trend/src/master/.

### 4.1 Trend-based and hybrid regulation toy models

The graph shown in the top left corner of Fig. 2 represents a simplified causal structure of intervention modeling scenarios, where intervention cannot target the root cause, but rather attenuates its downstream effects. The small graph has 4 nodes: `cause`, `intervention`, `problem`, and `outcome`. Fixing the input signal from `cause` and `intervention`, we are testing different combinations of their contributions, as well as different effects of `problem` on `outcome`.

We ran experiments for each of the level-based, trend-based, and hybrid regulation of `outcome` by `problem`, in 5 different scenarios as displayed in Table 1.

**Table 1: Toy model experiments on 4-element graph**

| # | Scenario name | Description |
|---|---|---|
| 1 | Regular | `problem` negatively regulates `outcome` |
| 2 | OR | `problem` and `cause` independently negatively regulate `outcome` |
| 3 | AND | `problem` and `cause` together negatively regulate `outcome` |
| 4 | NOT | `problem` positively regulates `outcome` |
| 5 | Target | `problem` and `cause` together negatively regulate `outcome` only when `cause`=0.7 |

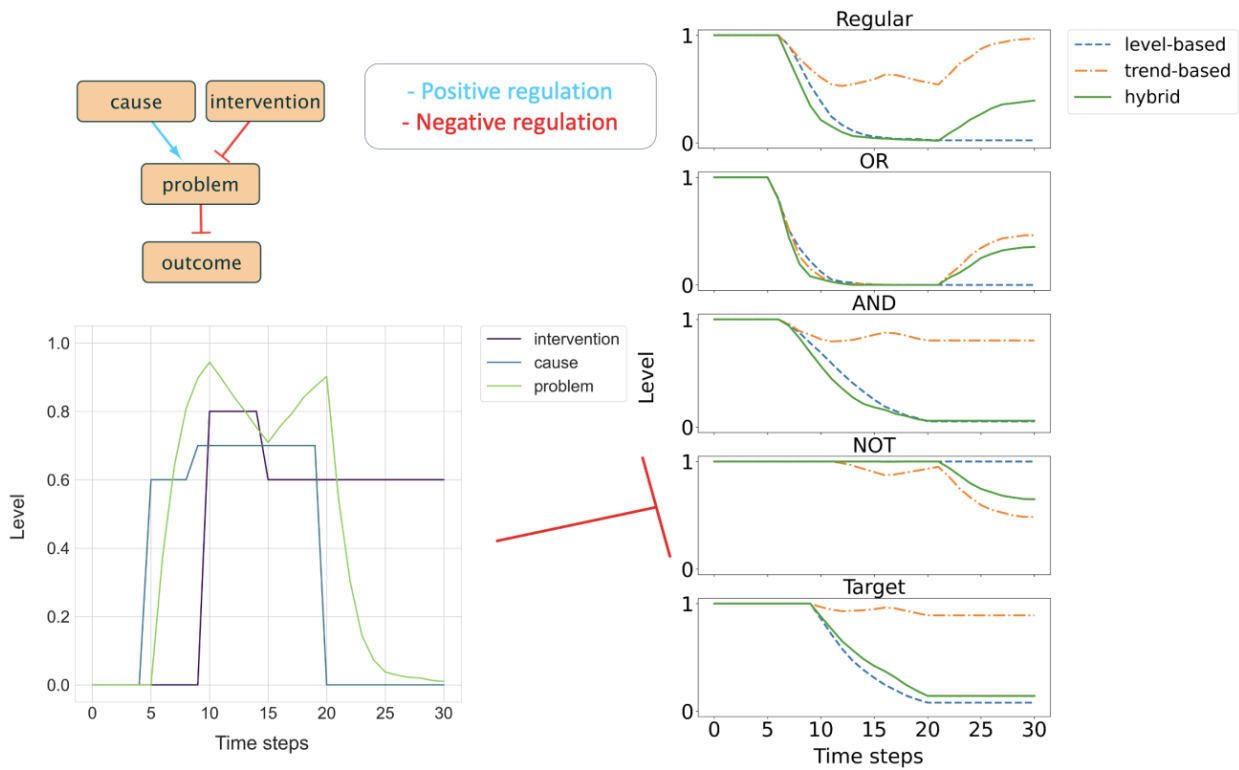

Figure 2: Toy model results: the underlying graph structure (top left); input signal on cause and intervention nodes, alongside their regulated element problem (bottom left); output comparison between level-based, trend-based, and hybrid regulation models in 5 different scenarios (right).

## 4.2 Modeling food insecurity in Ethiopia

We developed a causal model of food insecurity in the Ethiopian region Oromia as a use case for hybrid regulation. This model contains 31 nodes and 49 edges (including one self-edge), as shown in Fig. 3. The model is structured in such a way that it has input nodes for the causes of a potential crisis relevant to food security, `SARS-CoV-2 virus` denoted as `coronavirus` in Fig 3, `conflict`, and `locust swarm`, as well as intervention and aid nodes, `sanitation and hygiene aid` displayed as `sanitation aid`, and `food aid`. This structure facilitates hypothesis generation and comparisons between different intervention strategies.

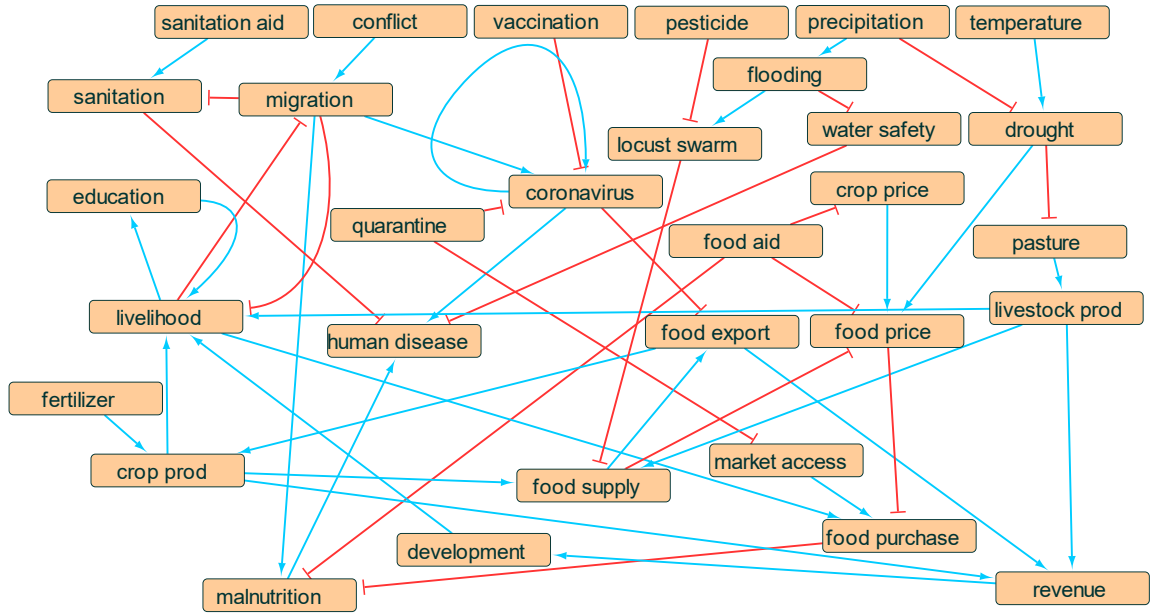

Figure 3: Large model of food security in Ethiopian region Oromia.

Simulation is set up with real-world data for most of the input nodes (Table 2) to explore the advantages of trend-based and hybrid regulation in modeling realistic scenarios. The biggest challenges were data scarcity and misalignment between time series data: some of the variables such as `temperature` and `precipitation` have daily resolution with data spanning over 70 years (Terraclimate, ERA5, CHIRPS), while `malnutrition` is monitored through surveys that only take place once in 4-5 years (DHS). Following available data, we modeled the period from January 2008 to March 2022, with monthly resolution (each simulation time step corresponds to one month) to have a resolution as fine-grained as possible, but limit interpolation as much as possible. Data given with daily resolution was aggregated by summing counts (`conflict`, `SARS-CoV-2 virus`) or averaging (`temperature`, `precipitation`). Data-derived input signals are shown in Fig. 4.

Each element is represented by an 11-level variable (with levels: {0, 0.1, 0.2, …1}), which allows convenient conversion between integer indices used to define value toggling and target regulation ({0, 1, 2, …, 10}), and variable values. Discretization is done by splitting the range of element's historical recorded values from available data into 11 uniformly distributed bins and then translating real number values into integer values of their bins (i.e. the lowest bin numbers into 0, the following bin into 1, etc.). Note that near the end of the simulation (after March 2021) there are no available data at the time when the simulations were run. For `temperature` and `precipitation`, we copied the last 12 months' data perpetually until the end of the simulation, because these are periodic, while for the other input variables we kept them at their last recorded value. The unknown variables are initialized at mid-level (0.5) in the absence of more accurate information. Note that, while we are treating `SARS-CoV-2 virus` as an input, in the model it is regulated by `quarantine` and `vaccination`. After March 2021 we are no longer enforcing its value via the value toggling feature, and its dynamics are predicted by DiSH-trend simulation.

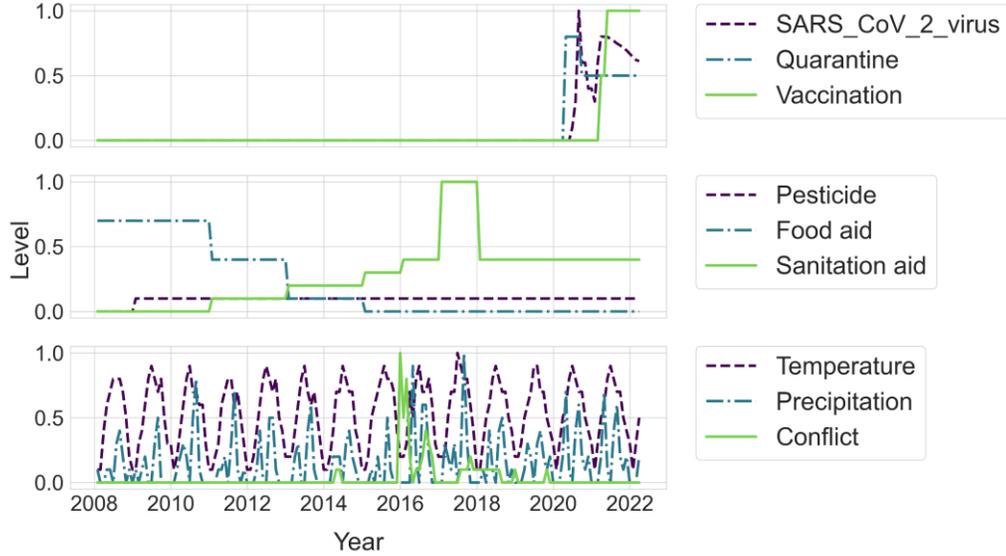

Figure 4: Data-derived input signals for the large food insecurity model.

While the simulator does not enforce the weights to be in the interval of [0,1], it is strongly recommended for simulation stability (avoiding saturation), and easier interpretability (absolute values of these weights can be interpreted as transition probabilities). With that in mind, we took value weights to be the maximum ones ($w_v = 1$) to observe the strongest level-based regulation effects. Consequently, for the trend-based regulation weights $w_t = 0.5$ to account for the fact that the trend can be negative. With the span of possible values [-1,1], the interval of influences [-0.5, 0.5], is the same size as level-based regulation interval of influences [0,1]. In hybrid regulation models, we set weights $w_v = 0.25$, and $w_t = 0.5$, to make it easier to compare them to the trend-based regulation (they would use the same trend influences, and the only difference would come from the level-based contributions).

Table 2: Data sources for Oromia food insecurity model with spatiotemporal aggregation level

| Variable | Source | Reference | Resolution (spatial/temporal) |
| --- | --- | --- | --- |
| Conflict | ACLED | Raleigh et al 2010 | Ethiopia/daily |
| Food aid | FAOSTAT | FAOSTAT | Ethiopia/yearly |
| Pesticide | FAOSTAT | FAOSTAT | Ethiopia/yearly |
| Precipitation | CHIRPS | Funk et al 2015 | Oromia/daily |
| Quarantine | EPHI | EPHI | Ethiopia/monthly |
| Sanitation and hygiene aid | USAID | USAID Database | Ethiopia/yearly |
| SARS-CoV-2 virus | OWID | OWID; Roser et al 2020 | Ethiopia/daily |
| Temperature | ERA5 | ERA5; Hersbach et al 2020 | Oromia/daily |
| Vaccination | WHO | WHO | Ethiopia/daily |

We run the simulation 200 times with 170 time steps and then average across the runs. The experiment is done three times: once for the level-based regulation model, once for the trend-based regulation, and once for the hybrid regulation model. We built the trend-based regulation model by manually selecting interactions where the regulated element value does not keep increasing/decreasing when its regulator stops increasing/decreasing and converted them to trend-based interactions. After that, we built the hybrid regulation model by converting all trend-based interactions into hybrid interactions, for comparison purposes.

## 5 RESULTS

The results show the advantages of DiSH-trend over DiSH in capturing element trends, as well as how DiSH-trend can recover qualitative behaviors missed by DiSH.

**5.1. Small network model results**

The simulated behavior of `outcome` in all five scenarios, for all three models, is displayed in Fig. 2. The regular model shows expected qualitative behavior: hybrid regulation has a steeper decrease than level-based regulation because the negative trend is contributing to it, as well. After time step 20, when the `problem` value decreases, its level-based influence is decreasing, but its negative trend is observed as a positive trend in the trend-based and hybrid regulated `outcome`, so these two show a progressively slowing increase.

OR scenario gives similar predictions as the Regular scenario for level-based and hybrid regulation, the only differences being undistinguishable trend-based and hybrid regulation. This happens because `cause` has a nonzero value between time steps 5 and 19, and it regulates `outcome` negatively and stronger than the trend contributions from `problem`. Only from time step 20, these trend contributions are not opposed by `cause`, and they increase the `outcome` value.

For the AND scenario in the level-based and the hybrid regulation, we observe a `problem`-induced decrease in `outcome`, but slower than seen in the Regular model, because these contributions are multiplied with 0.6, which is the value of `cause`. In trend-based regulation slower decrease means a smaller drop, so the `outcome` value saturates earlier.

The NOT scenario has the `outcome` update function with the same absolute value, but the opposite sign, compared to the Regular model. However, the element trajectories are not symmetric because the element values are bound in [0,1] intervals. Trend-based regulation cannot increase outcome value because it is already at maximum level, and the first change is observed when `problem` value increases, which in turn decreases `outcome` value. After that, the trend-based regulation `outcome` in the NOT scenario is symmetric to the trend-based regulation `outcome` in the Regular scenario. The level-based regulation `outcome` remains at maximum level throughout simulation because it starts at maximum level and the only regulator is `problem`, which increases `outcome` value.

In the Target scenario, `outcome` behaves similarly to the AND scenario, which is expected given that the Target scenario changes `outcome` value only when `cause` is at 0.7, and the AND scenario changes `outcome` value whenever `cause` is nonzero.

Small model simulation runtime is comparable between level-based, trend-based, and hybrid regulation, and was for all iterations between 0.374 s and 0.539 s.

**5.2 Large network model results**

The three versions of the model for food insecurity in Ethiopia, with level-based regulation only, trend-based regulation only, and hybrid regulation that combines value- and trend-based regulation, are examined by comparing predicted outcomes of non-fixed variables. A subset of variables with the most notable distinction is displayed in Fig. 5.

The resulting plots show that trend-based regulation modeling can predict different outcomes than traditional level-based regulation, ranging from moderate (`sanitation and hygiene`, and `human migration`) to opposite (`food supply`, `food export`, `livestock production`, and `human disease`). Evaluation of these outcomes requires more data, but some of the predicted outcomes can be qualitatively evaluated. There are examples where trend-based regulation yields more intuitive predictions than level-based regulation, such as `education` increasing, `human migration` increasing after a series of conflicts, and `sanitation and hygiene` increasing before conflicts but decreasing during conflicts and the pandemic. On the other hand, level-based predictions seem to favor food availability (`livestock production`, `food supply`) and expect `human migration` and `human diseases` to be progressively decreasing, which is expected in peaceful times. None of these regulation methods is better than the other, per se, but having them both increases predictive capacity by allowing more variety in reachable outcomes.

Hybrid regulation models strike a balance between value- and trend-based regulation models, by combining them, which is observed in the results. However, when a positive hybrid regulation is present, and the regulator value is increasing, then the regulated element's value will increase faster than with either value- or trend-based regulation because both value and trend are contributing to the increase (see `education` in Fig. 5), Hybrid regulation predicted outcomes can be similar to the ones predicted by level-based regulation, or trend-based regulation, depending on which is more prominent: regulator's current value or regulator's current trend. In this model, hybrid regulation outperformed level-based and trend-based regulation in capturing effects of a pandemic on the system: it captured the strongest rise in `human diseases`, and the deepest drop in `food export` and `food purchase`.

Large model simulation runtime on average for level-based regulation is 68.7 s, trend-based 71.2 s, and hybrid regulation 74.7 s.

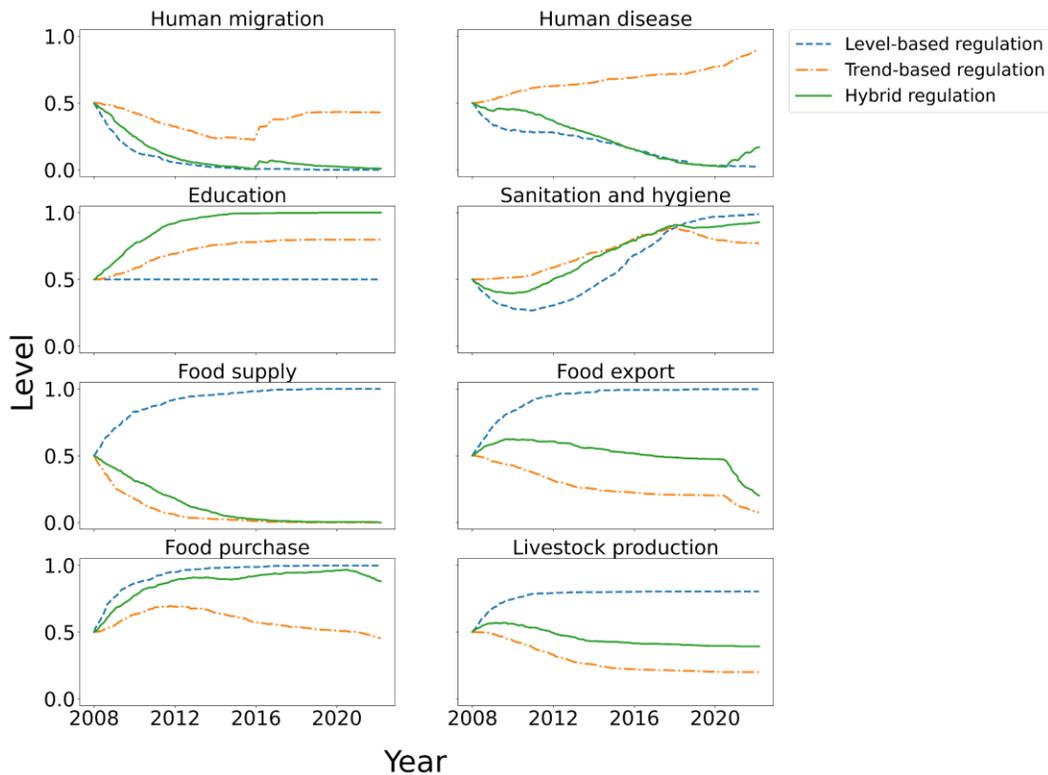

Figure 5: Comparison of selected simulation results between level-based (dashed blue line), trend-based (orange dot dash line) and hybrid regulation models (green full line).

## 6   CONCLUSIONS

In this paper, we present DiSH-trend, an extended version of the DiSH simulation tool for wide application in network dynamics modeling. DiSH simulator modeled dynamics by calculating updates for elements based on the values of their regulators. In DiSH-trend, these updates are extended to take into account regulator trends, as well. This allows for greater modeling flexibility and helps modelers simulate scenarios where the trend of the regulated element should follow the trend of the regulator.

We compared DiSH-trend with DiSH on a small 4-element toy model, and a larger 31-element model of food insecurity in Ethiopia. For each of the experiments, we considered level-based regulation (as DiSH extension which supports discrete elements), trend-based regulation, and hybrid regulation as a combination of the former two (DiSH-trend). The small model showed each of the implemented DiSH-trend functionalities, and how they yield distinct results between these three regulation modes. The large model showed how level-based regulation fails to capture some expected trends, and how hybrid regulation improves on that.

Allowing a greater variety of interactions by introducing trend-based regulation and combining it with traditional level-based regulation resulted in more faithful complex system representation in DiSH-trend, which in turn increased the simulator's predictive power. Leveraging more data and modeling at scenario-appropriate timescales will yield more conclusive and more precise predictions. The future work will focus on weight inference from data, which will increase the forecasting accuracy. Given data availability, integrating DiSH-trend with machine learning approaches for parameter tuning can leverage its speed and robustness in generating predictions, to make it an efficient self-trained artificial intelligence-based tool for studying complex systems.

## ACKNOWLEDGMENTS

This work is supported in part by DARPA award W911NF-18-1-0017 awarded to N. Miskov-Zivanov.

The authors would like to thank Kara Bocan for valuable insights on translation of real-world data into DiSH-trend variables, and Khaled Sayed for technical advice on DiSH simulator usage. Additionally, the authors are grateful to Brad Goodman for facilitating access to the historical data, as well as to Robyn Kozierok, Zoe Henschield, Lynette Hirschman, Pam Bhattacharya, and Pascale Proulx for discussions on real-world applications of trend-based modeling.

## REFERENCES

Abatzoglou, J. T., S. Z. Dobrowski, S. A. Parks, and K. C. Hegewisch. 2018. "TerraClimate, a high-resolution global dataset of monthly climate and climatic water balance from 1958–2015." *Scientific data*, *5*(1):1-12.

Albert, I., J. Thakar, S. Li, R. Zhang, and R. Albert. 2008. "Boolean network simulations for life scientists." *Source code for biology and medicine*, *3*(1):1-8.

Barrat, A., M. Barthélemy, R. Pastor-Satorras, and A. Vespignani. 2004. "The architecture of complex weighted networks." *Proceedings of the National Academy of Sciences of the United States of America*, *101*(11):3747–3752. https://doi.org/10.1073/pnas.0400087101

Chaouiya, C. 2007. "Petri net modelling of biological networks." *Briefings in bioinformatics*, *8*(4):210-219.

Chatfield, C. 1978. "The Holt‐Winters forecasting procedure." *Journal of the Royal Statistical Society*, *27*(3):264-279.

DHS, https://dhsprogram.com/data/, accessed on March 27th, 2021

ERA5, https://www.ecmwf.int/en/forecasts/datasets/reanalysis-datasets/era5, accessed on March 22nd, 2021

EPHI, https://www.ephi.gov.et/images/Registerd-COVID-19-Directive-2013_Final_051020.pdf, accessed on March 27th, 2021

Faeder, J. R., M. L. Blinov, B. Goldstein, and W. S. Hlavacek. 2005. "Rule-based modeling of biochemical networks." *Complexity*, *10*(4):22-41.

FAOSTAT, http://www.fao.org/faostat/en/#data, accessed on March 27th, 2021

Funk, C., P. Peterson, M. Landsfeld, D. Pedreros, J. Verdin, S. Shukla, G. Husak, J. Rowland, L. Harrison, A. Hoell, and Michaelsen, J. 2015. "The climate hazards infrared precipitation with stations—a new environmental record for monitoring extremes." *Scientific data*, *2*(1):1-21.

Hersbach, H., B. Bell, P. Berrisford, S. Hirahara, A. Horányi, J. Muñoz-Sabater, J. Nicolas, C. Peubey, R. Radu, D. Schepers, A. Simmons, C. Soci, S. Abdalla, X. Abellan, G. Balsamo, P. Bechtold, G. Biavati, J. Bidlot, M. Bonavita, G. De Chiara, P. Dahlgren, D. Dee, M. Diamantakis, R. Dragani, J. Flemming, R. Forbes, M. Fuentes, A. Geer, L. Haimberger, S. Healy, R. J. Hogan, E. Hólm, M. Janiskova, S. Keeley, P. Lalayoux, P. Lopez, C. Lupu, G. Radnoti, P. de Rosnay, I. Rozum, F. Vamborg. S. Villaume, and J. N. Thépaut. 2020. "The ERA5 global reanalysis." *Quarterly Journal of the Royal Meteorological Society*, *146*(730):1999-2049.

Holt, C. C. 2004. "Forecasting seasonals and trends by exponentially weighted moving averages." *International journal of forecasting*, 20(1):5-10.

Materi, W., & D. S. Wishart. 2007. "Computational systems biology in drug discovery and development: methods and applications." *Drug discovery today*, *12*(7-8):295-303.

Needham, C. J., J. R. Bradford, A. J. Bulpitt, and D. R. Westhead. 2007. "A primer on learning in Bayesian networks for computational biology." *PLoS Comput Biol*, *3*(8):e129.

OWID, https://github.com/owid/covid-19-data/tree/master/public/data, accessed on March 22nd, 2021

Raleigh, C., A. Linke, H. Hegre, and J. Karlsen. 2010. "Introducing ACLED: an armed conflict location and event dataset: special data feature." *Journal of peace research*, *47*(5):651-660.

Roser, M., H. Ritchie, E. Ortiz-Ospina, and J. Hasell. 2020. "Coronavirus pandemic (COVID-19)." *Our world in data*.

Sayed, K., Y. H. Kuo, A. Kulkarni, and N. Miskov-Zivanov. 2017. DiSH simulator: Capturing dynamics of cellular signaling with heterogeneous knowledge. In *2017 Winter Simulation Conference (WSC)*, IEEE:896-907.

USAID Database. 2021. https://explorer.usaid.gov/data, accessed on March 27th

WHO. 2021. "Ethiopia introduces COVID-19 vaccine national launching ceremony." https://www.afro.who.int/news/ethiopia-introduces-covid-19-vaccine-national-launching-ceremony, accessed on March 27th

Wilkinson, D. J. 2007. "Bayesian methods in bioinformatics and computational systems biology." *Briefings in bioinformatics*, *8*(2):109-116.

Winters, P. R. 1960. "Forecasting sales by exponentially weighted moving averages." *Management science*, 6(3):324-342.

**AUTHOR BIOGRAPHIES**

**STEFAN ANDJELKOVIC** is a PhD student in Joint Carnegie Mellon University and University of Pittsburgh PhD Program in Computational Biology. His email address is stefan.andjelkovic@pitt.edu

**NATASA MISKOV-ZIVANOV** is an Assistant Professor in Electrical and Computer Engineering, Bioengineering, and Computational and Systems Biology at the University of Pittsburgh. Her research interests include automated and rapid modeling of complex systems, knowledge-based reasoning in complex systems, systems and synthetic biology, and systems medicine. Her email address is nmzivanov@pitt.edu. Her website is https://www.nmzlab.pitt.edu.